\documentclass[english,pra,aps,amsmath,amssymb,showpacs,preprint]{revtex4}
\usepackage[T1]{fontenc}
\usepackage[latin9]{inputenc}
\usepackage{amsmath}
\usepackage{amssymb}
\usepackage{stmaryrd}
\usepackage{esint}

\makeatletter
\@ifundefined{textcolor}{}
{%
 \definecolor{BLACK}{gray}{0}
 \definecolor{WHITE}{gray}{1}
 \definecolor{RED}{rgb}{1,0,0}
 \definecolor{GREEN}{rgb}{0,1,0}
 \definecolor{BLUE}{rgb}{0,0,1}
 \definecolor{CYAN}{cmyk}{1,0,0,0}
 \definecolor{MAGENTA}{cmyk}{0,1,0,0}
 \definecolor{YELLOW}{cmyk}{0,0,1,0}
}

\usepackage{subfigure}\usepackage{hyperref}\usepackage{float}\hypersetup{
unicode=false,
pdftoolbar=true,
pdfmenubar=true,
pdffitwindow=false,
pdfstartview={FitH},
pdftitle={Mytitle},
pdfauthor={Author},
pdfsubject={Subject},
pdfcreator={Creator},
pdfproducer={Producer},
pdfkeywords={keywords},
pdfnewwindow=true,
colorlinks=true,
linkcolor=red,
citecolor=blue,
filecolor=magenta,
urlcolor=cyan
}

\makeatletter\newcommand{\Rmnum}[1]{\expandafter\@slowromancap\romannumeral #1@}\makeatother

\makeatother

\usepackage{babel}

\makeatother

\usepackage{babel}

\makeatother

\usepackage{babel}

\makeatother

\usepackage{babel}

\makeatother

\usepackage{babel}

\makeatother

\usepackage{babel}

\makeatother

\usepackage{babel}

\makeatother

\usepackage{babel}

\makeatother

\usepackage{babel}

\makeatother

\usepackage{babel}

\makeatother

\usepackage{babel}
\begin{document}

\title{Thermal effect on mixed state geometric phases for neutrino propagation
in a magnetic field}

\author{Da-Bao Yang}

\email{bobydbcn@163.com}

\affiliation{Department of fundamental physics, School of Science, Tianjin Polytechnic
University, Tianjin 300387, People's Republic of China}

\author{Ji-Xuan Hou}

\affiliation{Department of Physics, Southeast University, Nanjing 211189, People's
Republic of China}

\author{Ku Meng}

\affiliation{Department of fundamental physics, School of Science, Tianjin Polytechnic
University, Tianjin 300387, People's Republic of China}

\date{\today}
\begin{abstract}
In astrophysical environments, neutrinos may propagate over a long
distance in a magnetic field. In the presence of a rotating magentic
field, the neutrino spin can flip from left-handed neutrino to right-handed
neutrino. Smirnov demonstrated that the pure state geometric phase
due to the neutrino spin precession may cause resonantg spin conversion
inside the Sun. However, in general, the neutrinos may in an ensemble
of thermal state. In this article, the corresponding mixed state geometric
phases will be formulated, including the off-diagonal casse and diagonal
ones. The spefic features towards temperature will be analysized.
\end{abstract}


\maketitle

\section{Introduction}

\label{sec:introduction}

Geometric phase had been discovered by Berry \cite{berry1984quantal}
in the circumstance of adiabatic evolution. Then it was generalized
by Wilczek and Zee \cite{Wilczek1984appearance}, Aharonov and Anandan
\cite{aharonov1987phase,anandon1988nonadiabatic}, Samuel and Bhandari\cite{samuel1988general},
and Samuel and Bhandari \cite{samuel1988general} in the context of
pure state. Moreover, it was also extended to mixed state counterparts.
Its operationally well-defined notion was proposed by Sj$\ddot{o}$qvist
\emph{et. al}. \cite{sjoqvist2000mixed} based on inferferometry.
Subsequently, it was generalized to degenerate case by Singh et. al.
\cite{singh2003geometric} and to nonunitary evuolution by Tong et.
al. \cite{tong2004kinematic} by use of kenimatic approach. In addition,
when the final state is orthogonal to the initial state, the above
geometric phase is meaningless. So the complementary one to the usual
geometric phases had been put forward by Manini and Pistolesi \cite{manini2000off}.
The new phase is called off-diagonal geoemtric phase, which was been
generalized to non-abelian casse by Kult et. al. \cite{kult2007nonabelian}.
It also had been extended to mixed state ones by Filipp and Sj$\ddot{o}$qvist
\cite{filipp2003PRL,filipp2003offdiagonalPRA} during unitary evolution
. Further extension to non-degenerate case was made by Tong et. al.
\cite{tong2005offdiagonal} by kinematic approach. Finally , there
are excellent reviewed articles \cite{xiao2010Berry} and monographs
\cite{shapere1989book,bohm2003book,chru2004book} talking about its
influence and applications in physics and other natural science.

As well kown, Neutrino plays an important role in particle physics
and astronomy. Smirnov investigated the effect of resonant spin converstion
of solar neutrinos which was induced by the geometric phase \cite{smirnov1991solarneutrino}.
Joshi and Jain figured out the geometric phase of neutrino when it
was propagating in a rotating transverse mangnetic field \cite{joshi2015neutrino}.
However, their disscussion are confined to the pure state case. In
this article, we will talk about mixed state geometric phase of neutrino,
ranging from off-diagonal phase to diagonal one.

This paper is organised as follows. In the next section, the off-diagonal
geometric phase for mixed state will be reviewed as well as the usual
mixed state geometric phase. Furthermore, the related equation about
the propagation of two helicity components of neutrino will be retrospected.
In Sec. III, both the off-diagonal and diagonal mixed geometric phase
for neutrino in thermal state are going to be calculated. Finally,
a conclusion is drawn in the last section.

\section{Review of o ff-diagonal phase}

\label{sec:reviews}

If a non-degenerate density matrix takes this form
\begin{equation}
\rho_{1}=\lambda_{1}|\psi_{1}\rangle\langle\psi_{2}|+\cdots+\lambda_{N}|\psi_{N}\rangle\langle\psi_{N}|.\label{eq:DenstiyMatrix1}
\end{equation}
Moreover, a density operator that can't interfer with $\rho_{1}$
is introduced \cite{filipp2003offdiagonalPRA}, which is
\[
\rho_{n}=W^{n-1}\rho_{1}(W^{\dagger})^{n-1},n=1,...,N
\]
 where
\[
W=|\psi_{1}\rangle\langle\psi_{N}|+\psi_{N}\rangle\langle\psi_{N-1}|\cdots+|\psi_{2}\rangle\langle\psi_{1}|.
\]
In the unitary evolution, excet the ususal mixed state geometric phase,
there exists so called mixed state off diagonal phase , which reads
\cite{filipp2003offdiagonalPRA}
\begin{equation}
\gamma_{\rho_{j_{1}}...\rho_{j_{l}}}^{(l)}=\Phi[Tr(\prod_{a=1}^{l}U^{\parallel}(\tau)\sqrt[l]{\rho_{j_{a}}})],\label{eq:OffdiagonlaGeometricPhase}
\end{equation}
where $\Phi[z]\equiv z/|z|$ for nonzero complex number $z$ and \cite{tong2005offdiagonal}
\begin{equation}
U^{\parallel}=U(t)\sum_{k=1}^{N}e^{-i\delta_{k}},\label{eq:ParallelUnitaryEvolution}
\end{equation}
in which
\begin{equation}
\delta_{k}=-i\int_{0}^{t}\langle\psi_{k}|U^{\dagger}(t^{\prime})\dot{U}(t^{\prime})|\psi_{k}\rangle dt^{\prime}\label{eq:DynamicalPhase}
\end{equation}
and $U(t)$ is the time evolution operator of this system. Moreover
$U^{\parallel}$ satisfies the parallel transport condition, which
is
\[
\langle\psi_{k}|U^{\parallel\dagger}(t)\dot{U}^{\parallel}(t)|\psi_{k}\rangle=0,\ k=1,\cdots,N.
\]

In addition, the usual mixed state geometric phase factor \cite{tong2004kinematic}
takes the following form
\begin{equation}
\gamma=\Phi\left[\sum_{k=1}^{N}\lambda_{k}\langle\psi_{k}|U(\tau)|\psi_{k}\rangle e^{-i\delta_{k}}\right]\label{eq:DiagonalGeometricPhase}
\end{equation}

The propagation helicity components $\left(\begin{array}{cc}
\nu_{R} & \nu_{L}\end{array}\right)^{T}$of a neutrino in a magnetic field obeys the following equation \cite{smirnov1991solarneutrino}

\begin{equation}
i\frac{d}{dt}\left(\begin{array}{c}
\nu_{R}\\
\nu_{L}
\end{array}\right)=\left(\begin{array}{cc}
\frac{V}{2} & \mu_{\nu}Be^{-i\omega t}\\
\mu_{\nu}Be^{i\omega t} & -\frac{V}{2}
\end{array}\right)\left(\begin{array}{c}
\nu_{R}\\
\nu_{L}
\end{array}\right),\label{eq:Schrodinger}
\end{equation}
where $T$ denotes the matrices transposing operation, $\vec{B}=B_{x}+iB_{y}=Be^{i\omega t}$,
$\mu_{\nu}$ reprensents the magnetic moment of a massive Dirac neutrino
and $V$ is a term due to neutrino mass as well as interaction with
matter. The instantaneous eigenvalues and eigenvectors corresponding
to the Hamiltonian take the following form \cite{joshi2015neutrino}
\[
E_{1}=+\sqrt{\left(\frac{V}{2}\right)^{2}+(\mu_{\nu}B)^{2}}
\]
\begin{equation}
|\psi_{1}\rangle=\frac{1}{N}\left(\begin{array}{c}
\mu_{\nu}B\\
-e^{i\omega t}\left(\frac{V}{2}-E_{1}\right)
\end{array}\right)\label{eq:EigenVector1}
\end{equation}
and
\[
E_{2}=-\sqrt{\left(\frac{V}{2}\right)^{2}+(\mu_{\nu}B)^{2}}
\]
\begin{equation}
|\psi_{2}\rangle=\frac{1}{N}\left(\begin{array}{c}
e^{-i\omega t}\left(\frac{V}{2}-E_{1}\right)\\
\mu_{\nu}B
\end{array}\right),\label{eq:EigenVector2}
\end{equation}
where the normalized factor
\[
N=\sqrt{\left(\frac{V}{2}-E_{1}\right)^{2}+\left(\mu_{\nu}B\right)^{2}}.
\]
If this system in a thermal state, the density operator can be written
as
\begin{equation}
\rho=\lambda_{1}|1\rangle\langle1|+\lambda_{2}|2\rangle\langle2|\label{eq:DensityMatrix}
\end{equation}
where
\[
\lambda_{1}=\frac{e^{-\beta E_{1}}}{e^{-\beta E_{1}}+e^{-\beta E_{2}}}
\]
and
\[
\lambda_{2}=\frac{e^{-\beta E_{2}}}{e^{-\beta E_{1}}+e^{-\beta E_{2}}}.
\]
In addition, $\beta=1/(kT),$ where $k$ is the Boltzmann constant
and $T$ represents the temperature. In the next section, the mixed
state geometric phase for both off-diagonal one and diagonal one will
be calculated.

\section{Mixed state geometric phase}

\label{sec:Nonadiabatic}

The differential equation Eq. \eqref{eq:Schrodinger} can be exactly
solved by the following transformation
\begin{equation}
\left(\begin{array}{c}
\nu_{R}\\
\nu_{L}
\end{array}\right)=e^{-i\sigma_{z}\frac{1}{2}\omega t}\left(\begin{array}{c}
a\\
b
\end{array}\right),\label{eq:TransformedState}
\end{equation}
where $\sigma_{z}$ is a Pauli matrix along $z$ direction whose explicit
form is
\[
\sigma_{z}=\left(\begin{array}{cc}
1 & 0\\
0 & -1
\end{array}\right).
\]
By substituting Eq. \eqref{eq:TransformedState} into Eq. \eqref{eq:Schrodinger},
one can obtain
\begin{equation}
i\frac{d}{dt}\left(\begin{array}{c}
a\\
b
\end{array}\right)=\tilde{H}\left(\begin{array}{c}
a\\
b
\end{array}\right),\label{eq:TransformedSchodinger}
\end{equation}
where
\[
\tilde{H}=\mu_{\nu}B\sigma_{x}+\frac{1}{2}(V-\omega)\sigma_{z}.
\]
Furthermore, it can be written in this form
\begin{equation}
\tilde{H}=\frac{1}{2}\Omega\left(\begin{array}{ccc}
\frac{2\mu_{\nu}B}{\Omega} & 0 & \frac{V-\omega}{\Omega}\end{array}\right)\centerdot\left(\begin{array}{ccc}
\sigma_{x} & \sigma_{y} & \sigma_{z}\end{array}\right),\label{eq:TransformedHamiltonian}
\end{equation}
where $\Omega=\sqrt{\left(2\mu_{\nu}B\right)^{2}+\left(V-\omega\right)^{2}}$.
Because $\tilde{H}$ is independent of time, Eq. \ref{eq:TransformedSchodinger}
can be exactly solved, whose time evolution operator takes the form
\[
\tilde{U}=e^{-i\tilde{H}t}.
\]
Associating with Eq. \eqref{eq:TransformedState}, the time evolution
operator for Eq. \eqref{eq:Schrodinger} is
\begin{equation}
U=e^{-i\tilde{H}t}e^{i\sigma_{z}\frac{1}{2}\omega t}.\label{eq:UnitaryEvolution}
\end{equation}
By substituting Eq. \eqref{eq:TransformedHamiltonian} into Eq. \ref{eq:UnitaryEvolution},
the above operator can be written in an explicit form, which is
\[
U=\left(\begin{array}{cc}
\cos\frac{\Omega}{2}t-i\frac{V-\omega}{\Omega}\sin\frac{\Omega}{2}t & -i\frac{2\mu_{\nu}B}{\Omega}\sin\frac{\Omega}{2}t\\
-i\frac{2\mu_{\nu}B}{\Omega}\sin\frac{\Omega}{2}t & \cos\frac{\Omega}{2}t+i\frac{V-\omega}{\Omega}\sin\frac{\Omega}{2}t
\end{array}\right)\left(\begin{array}{cc}
e^{i\frac{\omega t}{2}} & 0\\
0 & e^{-i\frac{\omega t}{2}}
\end{array}\right)
\]
In order to calculate off-diagonal phase \eqref{eq:OffdiagonlaGeometricPhase},
by use of Eq. \eqref{eq:ParallelUnitaryEvolution}, we can work out
\[
\begin{array}{ccc}
U_{11}^{\parallel} & \equiv & \langle\psi_{1}|U(t)\left(e^{-i\delta_{1}}|\psi_{1}\rangle\langle\psi_{1}|+e^{-i\delta_{2}}|\psi_{2}\rangle\langle\psi_{2}|\right)|\psi_{1}\rangle\\
 & = & U_{11}e^{-i\delta_{1}},
\end{array}
\]
where $U_{11}=\langle\psi_{1}|U(t)|\psi_{1}\rangle$. In order to
simplify the result, let's talk about an easier case. when $t=\tau=2\pi/\Omega$,
\begin{equation}
U_{11}=-\frac{1}{N^{2}}\left[\mu_{\nu}^{2}B^{2}e^{i\frac{\omega\tau}{2}}+\left(\frac{V}{2}-E_{1}\right)^{2}e^{-i\frac{\omega\tau}{2}}\right]=U_{22}^{*},\label{eq:UDiagonal}
\end{equation}
where $*$ denotes the complex conjugate operation. By similar calculations,
one can obtains
\begin{equation}
U_{12}=\frac{2}{N^{2}}\mu_{\nu}B\left(\frac{V}{2}-E_{1}\right)\sin\left(\frac{1}{2}\omega\right)e^{-i(\omega\tau+\frac{\pi}{2})}=-U_{21}^{*}\label{eq:UOffDiagonal}
\end{equation}
Furthermore, $\delta_{1}$ can be explicitly calculated out by substituting
Eq. \eqref{eq:UnitaryEvolution} Eq. \eqref{eq:EigenVector1} into
Eq. \eqref{eq:DynamicalPhase}, which takes the form
\begin{equation}
\delta_{1}=\frac{1}{N^{2}}\left[2\mu_{\nu}^{2}B^{2}\left(\frac{V}{2}-E_{1}\right)+\left(\frac{V}{2}-E_{1}\right)^{2}\left(\frac{V}{2}-\omega\right)-\mu_{\nu}^{2}B^{2}\left(\frac{V}{2}-\omega\right)\right]\tau.\label{eq:DynamicalPhase1}
\end{equation}
 By similar calculation, one can get
\begin{equation}
\delta_{2}=-\delta_{1}.\label{eq:DynamicalPhase2}
\end{equation}
Hence Eq. \eqref{eq:ParallelUnitaryEvolution} can be explicitly calculated
out,
\begin{equation}
\left(\begin{array}{cc}
U_{11}^{\parallel} & U_{12}^{\parallel}\\
U_{21}^{\parallel} & U_{22}^{\parallel}
\end{array}\right)=\left(\begin{array}{cc}
U_{11} & U_{12}\\
U_{21} & U_{22}
\end{array}\right)\left(\begin{array}{cc}
e^{-i\delta_{1}} & 0\\
0 & e^{-i\delta_{2}}
\end{array}\right).\label{eq:RelationsUnitaryEvolutions}
\end{equation}

Now, let us calculate the mixed state off-diagonal phasse
\begin{equation}
\gamma_{\rho_{1}\rho_{2}}^{(2)}=\Phi\left[Tr\left(\prod_{a=1}^{2}U^{\parallel}(\tau)\sqrt{\rho_{a}}\right)\right],\label{eq:OffDiagonalGeometricPhaseForNeutrino}
\end{equation}
where $\rho_{1}=\lambda_{1}|1\rangle\langle1|+\lambda_{2}|2\rangle\langle2|$
and $\rho_{2}=\lambda_{1}|2\rangle\langle2|+\lambda_{2}|1\rangle\langle1|$.
Under the basis of $|\psi_{1}\rangle$ and $|\psi_{2}\rangle$,
\begin{equation}
\begin{array}{ccc}
Tr\left(\prod_{a=1}^{2}U^{\parallel}(\tau)\sqrt{\rho_{a}}\right) & = & \sum_{b=1}^{2}\langle\psi_{b}|\prod_{a=1}^{2}U^{\parallel}(\tau)\sqrt{\rho_{a}}|\psi_{b}\rangle\\
 & = & \sqrt{\lambda_{1}\lambda_{2}}\left[\left(U_{11}^{\parallel}\right)^{2}+\left(U_{22}^{\parallel}\right)^{2}\right]+U_{12}^{\parallel}U_{21}^{\parallel}.
\end{array}\label{eq:TraceOffDiagonal}
\end{equation}
By substituting Eq. \eqref{eq:RelationsUnitaryEvolutions} into Eq.
\eqref{eq:TraceOffDiagonal}, we can obtain a simpler result
\[
Tr\left(\prod_{a=1}^{2}U^{\parallel}(\tau)\sqrt{\rho_{a}}\right)=\sqrt{\lambda_{1}\lambda_{2}}\left[\left(U_{11}e^{-i\delta_{1}}\right)^{2}+\left(U_{22}e^{-i\delta_{2}}\right)^{2}\right]+U_{12}U_{21}e^{-i\left(\delta_{1}+\delta_{2}\right)}
\]
By substituting Eq. \eqref{eq:UDiagonal} and Eq. \eqref{eq:UOffDiagonal}
into the above equation, off-diagonal geometric phase \eqref{eq:OffDiagonalGeometricPhaseForNeutrino}
can be explicitly calculated,
\begin{equation}
\begin{array}{ccc}
\gamma_{\rho_{1}\rho_{2}}^{(2)} & = & \Phi\{\left(\frac{V}{2}-E_{1}\right)^{2}\mu_{\nu}^{2}B^{2}\left(\cos\omega\tau-1\right)+\sqrt{\lambda_{1}\lambda_{2}}\vartimes\\
 &  & [\left(\frac{V}{2}-E_{1}\right)^{4}\cos\left(\omega\tau+2\delta_{1}\right)+\mu_{\nu}^{4}B^{4}\cos\left(\omega\tau+2\delta_{1}\right)\\
 &  & +2\mu_{\nu}^{2}B^{2}\left(\frac{V}{2}-E_{1}\right)^{2}\cos2\delta_{1}]\}
\end{array}\label{eq:OffDiagonalPhaseFinalResult}
\end{equation}
Hence, the corresponding phase is either $\pi$ or $0$, which depends
on temperature and magnetic field. So, its phase is unresponsitive
to temperature.

By substituting Eq. \eqref{eq:UDiagonal}, Eq. \eqref{eq:DynamicalPhase1}
and Eq. \eqref{eq:DynamicalPhase2} into Eq. \eqref{eq:DiagonalGeometricPhase},
the diagonal geometric phase for miexed state reads
\begin{equation}
\begin{array}{ccc}
\gamma & = & \Phi\{\left[\lambda_{1}e^{i(\frac{\omega\tau}{2}-\delta_{1})}+\lambda_{2}e^{-i(\frac{\omega\tau}{2}-\delta_{1})}\right]\mu_{\nu}^{2}B^{2}+\\
 &  & \left[\lambda_{1}e^{-i(\frac{\omega\tau}{2}+\delta_{1})}+\lambda_{2}e^{i(\frac{\omega\tau}{2}+\delta_{1})}\right]\left(\frac{V}{2}-E_{1}\right)^{2}\}.
\end{array}\label{eq:DiagonalPhaseFinalResult}
\end{equation}
From the above result, we can draw a conclution that if $\lambda_{1}=\lambda_{2}$,
in another word $T\rightarrow\infty$, the corresponding phase maybe
$\pi$ or $0$. In other circumstance, it may vary continuously in
an interval. By contrary to off-diagonal one, the diagonal phase is
more sensitive to temprature.

\section{Conclusions and Acknowledgements }

\label{sec:discussion}

In this article, the time evolution operator of neutrino spin in the
presence of uniformly rotating magnetic field is obtained. Under this
time evolution operator, a thermal state of this neutrinos evolves.
Then there exists mixed off-diagonal geometric phase for mixed state,
as well as diagonal ones. They have been calculated respectively.
And an analytic form is achieved. In addition, a conclusion is drawn
that diagonal phase is more sensentive to off-diagonal one towards
temperature.

D.B.Y. is supported by NSF ( Natural Science Foundation ) of China
under Grant No. 11447196. J.X.H. is supported by the NSF of China
under Grant 11304037, the NSF of Jiangsu Province, China under Grant
BK20130604, as well as the Ph.D. Programs Foundation of Ministry of
Education of China under Grant 20130092120041. And K. M. is supported
by NSF of China under grant No.11447153.


\begin{thebibliography}{10}

\bibitem{anandon1988nonadiabatic}
J~Anandan.
\newblock Non-adiabatic non-abelian geometric phase.
\newblock {\em Physics Letters A}, 133(4-5):171--175, 1988.

\bibitem{berry1984quantal}
M.~V. Berry.
\newblock Quantal phase factors accompanying adiabatic changes.
\newblock {\em Proceedings of the Royal Society of London. Series A,
  Mathematical and Physical Sciences}, 392:45--57, 1984.

\bibitem{bohm2003book}
A~Bohm, A~Mostafazadeh, H~Koizumi, Q~Niu, and J~Zwanziger.
\newblock {\em The geometric phase in quantum systems}.
\newblock Springer-Verlag, 2003.

\bibitem{chru2004book}
D.~Chruscinski and A.~Jamioikowski.
\newblock {\em Geometric phases in classical and quantum mechanics}, volume~36.
\newblock Birkhauser, 2004.

\bibitem{filipp2003offdiagonalPRA}
Stefan Filipp and Erik Sjoqvist.
\newblock Off-diagonal generalization of the mixed-state geometric phase.
\newblock {\em Physical Review A}, 68(4):042112, 2003.
\newblock Copyright (C) 2010 The American Physical Society Please report any
  problems to prola@aps.org PRA.

\bibitem{filipp2003PRL}
Stefan Filipp and Erik Sjoqvist.
\newblock Off-diagonal geometric phase for mixed states.
\newblock {\em Physical Review Letters}, 90(5):050403, 2003.
\newblock Copyright (C) 2010 The American Physical Society Please report any
  problems to prola@aps.org PRL.

\bibitem{joshi2015neutrino}
Sandeep Joshi and Sudhir~R Jain.
\newblock Geometric phase for neutrino propagation in a transverse magnetic
  field.
\newblock 2015.

\bibitem{kult2007nonabelian}
D~Kult.
\newblock Non-abelian generalization of off-diagonal geometric phases.
\newblock {\em EPL (Europhysics Letters)}, 78:60004, 2007.

\bibitem{manini2000off}
Nicola Manini and F.~Pistolesi.
\newblock Off-diagonal geometric phases.
\newblock {\em Physical Review Letters}, 85(15):3067, 2000.
\newblock Copyright (C) 2010 The American Physical Society Please report any
  problems to prola@aps.org PRL.

\bibitem{samuel1988general}
Joseph Samuel and Rajendra Bhandari.
\newblock General setting for berry's phase.
\newblock {\em Physical Review Letters}, 60(Copyright (C) 2010 The American
  Physical Society):2339, 1988.
\newblock PRL.

\bibitem{shapere1989book}
A~Shapere and F~Wilczek.
\newblock Geometric phases in physics.
\newblock 1989.

\bibitem{singh2003geometric}
K.~Singh, D.~M. Tong, K.~Basu, J.~L. Chen, and J.~F. Du.
\newblock Geometric phases for nondegenerate and degenerate mixed states.
\newblock {\em Physical Review A}, 67(3):032106, 2003.
\newblock Copyright (C) 2011 The American Physical Society Please report any
  problems to prola@aps.org PRA.

\bibitem{sjoqvist2000mixed}
Erik Sjoqvist, Arun~K. Pati, Artur Ekert, Jeeva~S. Anandan, Marie Ericsson,
  Daniel K.~L. Oi, and Vlatko Vedral.
\newblock Geometric phases for mixed states in interferometry.
\newblock {\em Physical Review Letters}, 85(14):2845, 2000.
\newblock Copyright (C) 2010 The American Physical Society Please report any
  problems to prola@aps.org PRL.

\bibitem{smirnov1991solarneutrino}
A~Yu Smirnov.
\newblock The geometrical phase in neutrino spin precession and the solar
  neutrino problem.
\newblock {\em Physics Letters B}, 260(1):161--164, 1991.

\bibitem{tong2004kinematic}
D.~M. Tong, E.~Sjoqvist, L.~C. Kwek, and C.~H. Oh.
\newblock Kinematic approach to the mixed state geometric phase in nonunitary
  evolution.
\newblock {\em Physical Review Letters}, 93(8):080405, 2004.
\newblock Copyright (C) 2010 The American Physical Society Please report any
  problems to prola@aps.org PRL.

\bibitem{tong2005offdiagonal}
D.~M. Tong, Erik Sjoqvist, Stefan Filipp, L.~C. Kwek, and C.~H. Oh.
\newblock Kinematic approach to off-diagonal geometric phases of nondegenerate
  and degenerate mixed states.
\newblock {\em Physical Review A}, 71(3):032106, 2005.
\newblock Copyright (C) 2011 The American Physical Society Please report any
  problems to prola@aps.org PRA.

\bibitem{Wilczek1984appearance}
F~Wilczek and A~Zee.
\newblock Appearance of gauge structure in simple dynamical systems.
\newblock {\em Physical Review Letters}, 52(24):2111--2114, 1984.

\bibitem{xiao2010Berry}
Di~Xiao, Ming-Che Chang, and Qian Niu.
\newblock Berry phase effects on electronic properties.
\newblock {\em Reviews of Modern Physics}, 82(3):1959--2007, 2010.
\newblock RMP.

\bibitem{aharonov1987phase}
A.~Anandan Y.Aharonov.
\newblock Phase change during a cyclic quantum evolution.
\newblock {\em Phys. Rev. Lett}, 58(1593), 1987.

\end{thebibliography}
\end{document}